# Political Polarization in Social Media: Analysis of the "Twitter Political Field" in Japan

Hiroki Takikawa
Frontier Research Institute for Interdisciplinary Sciences
Tohoku University
Miyagi, Japan
takikawa@m.tohoku.ac.jp

Kikuko Nagayoshi
Graduate School of Arts and Letters
Miyagi, Japan
Tohoku University
nagayoshi@m.tohoku.ac.jp

*Abstract*—There is an ongoing debate about whether the Internet is like a public sphere or an echo chamber. Among many forms of social media, Twitter is one of the most crucial online places for political debate. Most of the previous studies focus on the formal structure of the Twitter political field, such as its homophilic tendency, or otherwise limits the analysis to a few topics. In order to explore whether Twitter functions as an echo chamber in general, however, we have to investigate not only the structure but also the contents of Twitter's political field. Accordingly, we conducted both large-scale social network analysis and natural language processing. We firstly applied a community detection method to the reciprocal following network in Twitter and found five politically distinct communities in the field. We further examined dominant topics discussed therein by employing a topic model in analyzing the content of the tweets, and we found that a topic related to xenophobia is circulated solely in right-wing communities. To our knowledge, this is the first study to address echo chambers in Japanese Twitter political field and to examine the formal structure and the contents of tweets with the combination of large-scale social network analysis and natural language processing.

## I. INTRODUCTION

There is an ongoing debate about whether the Internet promotes public dialogue, thus enlarging the public sphere, or exacerbates the tendency for like-minded people to group together, thus creating echo chambers (for optimistic arguments, see [7][10]; for pessimistic arguments, see [11][22]). Social media, which are sites where ordinary people generate, post, and transmit their own contents, including political opinions and ideas, are the main research sites for the "public sphere or echo chamber" question. Among numerous social media platforms, Twitter is the one on which many previous studies have focused [2][5][8][9][12][14].

Many studies have examined the extent of homophilic tendencies in following networks in Twitter. This is because homophilic ties are considered to be "building blocks of echo chambers" [5]. In other words, those previous studies addressed the (possible) formal structure of echo chambers. However, the idea of echo chambers itself refers to not only the structure but also the content of political speech and dialogue. It stipulates that, in echo chambers, similar opinions are circulated repeatedly, leading to the reinforcement of existing opinions and the fragmentation of speech communities. Although some studies have addressed the contents of tweets, they only examined a narrow range of particular predefined topics that are reduced to a binary value created by human coding. In order to fully understand the mechanism of echo chambers, however, it is necessary to analyze in depth both the structure and broad contents of speeches circulated in the network. Here, by employing a combination of large-scale social network analysis and natural language processing, we investigated the structure and the contents of Twitter's political field in Japan at the same time. This will shed new light on the relationship between the Internet and the public sphere.

## II. BACKGROUND

Political polarization and fragmentation is a serious danger to a well-functioning democracy [22]. According to J.S. Mill's conception of political philosophy, democracy needs a certain level of diversity in which citizens are exposed to a wide variety of ideas and opinions from fellow citizens and are willing to treat others as respected citizens and engage public dialogue with them.

Echo chambers created by the Internet would undermine this possibility in a serious way. In an echo chamber situation, people may regard those holding different opinions as enemies and retreat into segregated communities. If social media in the Internet creates this fragmentation, it would endanger the foundation of our democratic societies. It becomes, thus, an urgent question whether social media promotes public sphere or creates echo chambers. We tackle this question by investigating the structure and the content of Twitter for politically engaged people in Japan.

Twitter is a social networking service providing users the ability to send texts of a maximum of 140 characters to their followers and read texts from users they follow on Twitter (see [18]). In fact, Twitter users can follow any other users without explicit permission insofar as the accounts are not protected. Because of this characteristic, Twitter has more potential to be an open public space and well-suited for public dialogue than other social media, such as Facebook.

Twitter has not only played an important role in political discussion online but also on "real world" politics. It is well-known that during Arab Spring, Twitter was an effective tool for diffusing information and mobilizing people to the movements [21]. Recent social movements in various countries—such as the Spanish Indignados movements, the Occupy movement in the United States, and the Gezi movement in Turkey—have also employed Twitter as the instrumental means of communication [3][6]. On the other hand, Twitter also has been the subject of harsh criticisms for failing to manage hate speech and prevent the dissemination of fake news and misinformation.

In Japan, SNS is also broadly used. According to a survey conducted in 2015, around 31% of the respondents use Twitter [19]. The proportion is slightly smaller than Facebook users (35.3%). However, among the younger populations (those in their teens and twenties), 50% are Twitter users, surpassing the percentage of Facebook users.

One of the features of Twitter is its anonymity. The same survey shows that around 75% of Twitter users remain anonymous in 2014 in Japan, which is more than twice than in the U.S., U.K., or Korea [20]. Anonymity can promote deliberative democracy because it enables minority populations to express their opinions without concern of being identified. In Japan, however, anonymity seems to harm democracy because it allows citizens to openly express their prejudice. SNS, especially Twitter, is sometimes raised as a reason that hate speech against foreign citizens has spread in recent decades [23][26].

Internet is an effective tool for those involved in radical right-wing movements because it enables them to recruit new members cheaply and effectively [1]. Specifically, SNS presents one of "the most ideal platforms for sharing their own information, views, and communities with the masses in a mainstream venue virtually free of gatekeepers and oversight" [16]. In Japan as well, online activists of Japanese alt-right, called *Netto Uyoku*, has become one of the major social issues since the 2000s [24]. They are extreme nationalists who exhibit xenophobic attitudes toward China and North and South Korea in the cyber space, such as at online board *2 channel* and Twitter. Tsuji estimated that less than one percent of the population belong to *Netto Uyoku*, using results of an online survey [24]. Although the movement may have few followers at this time, their activities seem to have a strong impact. Many young members of the radical right-wing groups came to know the "hidden fact," such as conspiracy theories about China and Korea, through the Internet and are thus motivated to join in the group [13]. According to the results from an online survey, more than one third of *Netto Uyoku* has Twitter accounts and use Twitter every day, which is more active and widespread usage compared to the general population [24]. The same survey found no differences between *Netto Uyoku* and the other respondents in the frequency of Facebook or Line usage, suggesting that radical right-wing groups in Japan have an affinity for Twitter.

Is this affinity as a result of an echo chamber on Twitter that helps radicalize the attitudes of online activists? They legitimate their perspectives by associating different kinds of information such as news and political blogs. Through this web, hate-based knowledge becomes legitimate and is turned into web-based knowledge [16]. In Japan, hate-based tweets tend to be linked to the online board *2 channel* as a source of information. It has been proved that having contact with *2 channel* strengthens racist attitudes [23][25]. Thus, Twitter seems to work as a tool of expanding "hate-based knowledge" in *2 channel* to the public. If Twitter users connect only with those with similar political orientations, this may enhance the probability for them to accept the "hate-based knowledge" because they have few chances to examine the validity of the knowledge from different viewpoints. No studies exist that examine network structure and content of tweet at the same time. In light of this, it is worthwhile to explore whether Twitter functions as a public sphere or an echo chamber.

### III. PREVIOUS STUDIES

We classified previous studies into two types according to data collection and the characteristics of analysis.

The first type is the "entire field approach." These studies are typically interested in the formal structure of the entire political field in Twitter and investigate whether it promotes echo chambers or public dialogue. These studies identify the entire political field in Twitter in the following way [2][5][8][12]. First, they search for political actors such as national politician, political activists, and media outlets in Twitter. Using their account information, they extract their followers' lists corresponding to the list of politically engaged Twitter users. Based on this, they construct the entire field of the following network of politically engaged users. The advantage of this approach is that it covers as many politically engaged users as possible without the restriction of predefined topics or keywords.

The political orientation for each user is inferred from the "following" pattern of each user or the contents of their tweets. In the former case, for example, users who follow Republican politician in the U.S. are inferred to have conservative orientation [5][8][12]. In a more complex case, the ideal point estimation is used on the basis of the information of all accounts the users follow [2]. In the latter case, the contents of tweets are used to identify users' political orientation with a supervised machine-learning technique [8].

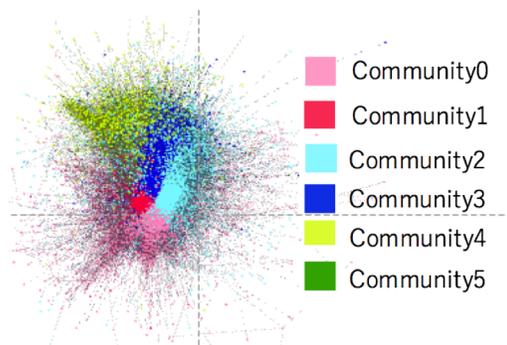

Figure 1. Six Distinct Speech Communities in Japanese Twitter Political Field

Then, researchers examine to what extent political homophily operates in the Twitter political field. Political homophily here refers to the tendency of people linking with those with similar political orientation. Homophily is considered to be the main cause of creating the structure of echo chambers. The basic results converge on the conclusion that there is a homophilic tendency in Twitter. The detailed results are, however, mixed. For example, Boutyline and Willer reported the conservatives are more homophilic than liberals [5], while Colleoni and colleagues showed the opposite result if each user's political orientation is identified from tweet contents, not just following behavior [8]. In any case, the main focus of Colleoni et al. is the homophilic structure of twitter users, not the contents of political discourse itself.

The second type of previous studies focuses on restricted samples or singular topics. We call it the "selective approach." These studies first search for the relevant users by using keywords or hashtag search [9][14]. For example, the users are identified as political engaged if they use hashtags such as "#democrats" or "#obama" [9], or alternatively use keywords in their tweets such as "Tea Party" or "Unemployment Benefits" [14]. Next, the network among users is constructed from following behaviors, Retweet (RT) behaviors, or mention behaviors. Finally, a cluster analysis or a community detection method is applied to these constructed networks. These clustering methods are used to examine whether users in their samples are segregated into different communities or integrated into public sphere.

This second type of studies take into consideration not only the structure but also the content of tweets, in contrast with the "entire field approach." Conover and colleagues hand-coded the tweets as liberal or conservative and found one-sided argument (either liberal or conservative) in each community in RT network [9]. Himelboim and colleagues took the similar procedure to demonstrate political homogeneity within communities in their sample [14]. Although their investigation of the contents of tweets would be one valuable step in investigating the mechanism of political polarization or echo chambers, the reduction of the contents to the binary "left" and "right" may hinder us from investigating its nuances. For example, it prevents us from asking what topics drive polarization or how differently distributed topics are among different communities.

In sum, in terms of data collection, the entire field approach would be better than the selection approach because it covers politically engaged people in a given area more effectively. However, previous studies of this approach seldom considered the content of political tweets. On the other hand, the studies addressing the content of speech usually take the restricted or selective approach to collect the relevant data. Furthermore, the tweet contents they examine are severely reduced to the simple binary value based on predetermined coding scheme.

Here we adopted the entire field approach and at the same time analyze the rich content of the tweets. Furthermore, we tried to avoid reducing the content of tweets to simple values (left–right). Specifically, we applied topic model to the contents of tweets, which allowed us to preserve the contents of tweets as much as possible while making them suitable for quantitative analysis.

Our research questions are as follows.

First, is the entire political field in Japanese Twitter differentiated into distinct political speech communities or does it rather have the integrative structure? In order to answer this question, we adopted the inductive method, that is, we applied a community detection method to the field, enabling us to inductively find the divided lines in the field. This method is more suitable for Japanese political field because, unlike the U.S. and other countries, it may not have clear-cut left–right division. Thus, this approach offers the potential to reveal multiple divided lines that are difficult to define a priori.

Second, if there are different communities in Twitter's political field, do those communities function as echo chambers or do they frequently communicate with each other? We examined the extent of echo chambers by looking at how much difference there are in terms of topics discussed by members. If a certain topic is discussed in exclusively a sole community, it would be concluded that the community functions as an echo chamber. On the other hand, if several topics are discussed to equal or similar extent among different communities, polarization is not as serious even if there are different speech communities in the Twitter political field.

IV. DATA

The data collection was done through Twitter REST API, from March 2017 through May 2017.
We adopted the entire field approach in collecting Twitter data. We tried to cover the Japanese Twitter political field as much as possible, which is defined as the sphere where politically engaged people exchange their ideas and opinions. Specifically, we gathered the data in the following way. First, we searched for the accounts of political party leaders in Japanese national politics in April 2017 (as for the "Nippon no Kokoro" party, we used the party official account because the leader has no personal account). Nine accounts of eight political parties (Liberal Party had two joint representative) were found. Then, by using the information on followers for each political leader, we identified politically engaged people as those who followed at least one of the political party leaders. The information regarding which political leaders each user follows, in turn, was used to specify his or her political orientation. Furthermore, we focused on the followers of the political leaders whose Twitter accounts were unprotected, had more than 500 followers, tweeted more than 1000 times, and posted at least one tweet in 2017. The total number of users was 92,313. This restriction was necessary for data manageability but also had a theoretical justification. By doing this, we can exclude bots and inactive users and focus on moderate opinion leaders who share political information, exchange political opinions, and engage in political dialogue.

Next, we gathered followers' data, including screen name, friends/followers list, profile description, and other information, through Twitter REST API (hereinafter "user data"). The friends/followers lists were used to construct the following network among the users in the Japanese Twitter political field, and the profile description was employed to identify users' political, social, and cultural orientation.

We also collected the tweet content data, which consisted of every user's latest tweets (due to API's constraints, the maximum number of tweets we could collect were the last 3200 tweets). This text data was preprocessed and used as the corpus for topic modeling.

## V. METHOD

Regarding network construction, we focused on a kind of reciprocal following network. Generally, following relation is not necessarily reciprocated in Twitter. Users are allowed to follow other users in one-way direction without permission. The difference between one-way following network and reciprocal following network in Twitter, as a previous study suggests, is that the former reflects the characteristics of a "newsy" media while reciprocal following networks better captures the aspect of social media in Twitter [18]. Since we are mainly interested in whether Twitter as a social media functions as public sphere, we restricted our analysis to a reciprocal following network in the following analysis.

The reciprocal following network was constructed based on user data, which contains follower/friends (following) lists. For example, the edge between user A and user B should be added if user A followed user B and user B also followed user A.

Then, we applied The Louvain method—one type of modularity-based community detection methods—to this network in order to obtain community partition in the Twitter political field. Modularity measures the density of edges within communities compared to edges between communities. The Louvain method is a fast heuristic algorithm for finding community partition to optimize modularity in a large network. Sociologically, each community detected by the method corresponds to a relatively cohesive speech community in the field. We used the NetworkX library in Python for the implementation of community detection algorithm.

Next, we applied Latent Dirichlet Allocation (LDA) [4], to the tweet content data. LDA is a standard type of topic models. Here each document is allowed to have mixture of topics and each topic, which means a set of probability distribution over vocabularies in the corpus, is assigned to each document probabilistically. Thus, each document is characterized by the topic proportion assigned to it.

Before implementing a topic model, we preprocessed the tweet data by applying morphological analysis to identify the unit of Japanese words and removing stop words. We also converted the texts into 1-gram bag of words. For documents, we merged all tweets posted by the same user into one document because one tweet contains only a maximum of 140 characters, making it difficult to appropriately assign it to a particular topic.

It has been reported that pooling all tweets by the same user into one document improves the precision of topic assignment [15]. This yielded the complete data sample: 79,536 documents, 125,745 vocabularies, and 402,640,579 tokens. After these preprocessing, we applied LDA to the data and implemented parallelized Latent Dirichlet Allocation by genism in Python library.

TABLE I. SIZE AND FOLLOW RATIO FOR EACH PARTY LEADER OF COMMUNITIES

| Community | Size | Follow ratio | | | | | | | | |
|---|---|---|---|---|---|---|---|---|---|---|
| | | Shinzo Abe (LDP) | Ichiro Matsui (JRP) | Mizuho Fukushima (SDP) | Nippon no Kokoro | Ichiro Ozawa (LP) | Renho (DPJ) | Kazuo Shii (JCP) | Natsuo Yamaguchi (KOMEI) | Taro Yamamoto (LP) |
| 0: Abe (LDP) follower | 16,559 | 0.74 | 0.06 | 0.08 | 0.05 | 0.06 | 0.22 | 0.04 | 0.01 | 0.09 |
| 1: Right-wing follower | 8,871 | 0.87 | 0.20 | 0.12 | 0.53 | 0.07 | 0.13 | 0.06 | 0.02 | 0.07 |
| 2: Renho (DPJ) follower | 40,241 | 0.37 | 0.07 | 0.20 | 0.02 | 0.14 | 0.50 | 0.07 | 0.01 | 0.27 |
| 3: Left-wing follower | 12,012 | 0.16 | 0.05 | 0.54 | 0.02 | 0.48 | 0.32 | 0.43 | 0.02 | 0.66 |
| 4: Yamamoto follower | 4,998 | 0.23 | 0.03 | 0.11 | 0.01 | 0.10 | 0.17 | 0.11 | 0.00 | 0.58 |
| 5: JRP and ruling party | 606 | 0.68 | 0.63 | 0.12 | 0.09 | 0.12 | 0.21 | 0.08 | 0.23 | 0.09 |
| Base line follow ratio | | 0.49 | 0.08 | 0.20 | 0.07 | 0.16 | 0.34 | 0.11 | 0.01 | 0.27 |

TABLE II. TOPIC DISTRIBUTIONS FOR EACH COMMUNITY

| Topic | 12: Idol | 24: Anime | 34: Xenophobia | 17: Cuisine | 40: Conspiracy law, Moritomo Gakuen scandal | 29: Music | 36: the mixed topic (Nuclear power plant and JRP) |
|---|---|---|---|---|---|---|---|
| Associated words | 生誕祭 (Birthday party)<br>うれし涙(Happy tears)<br>泣き顔(Crying face)<br>好きな人 (A person I love)<br>音符 (Music note) | 入荷 (Arrival of goods)<br>けものフレンズ (Title of Anime)<br>艦これ(Title of Anime)<br>抽選(Lot)<br>送料(Mailing cost) | 民主党 (DPJ)<br>蓮舫(Renho)<br>在日 (Korean Japanese)<br>韓国人(Korean)<br>反日(Anti-Japan) | ワイン (Wine)<br>講座 (Seminar)<br>香り (Aroma)<br>ご飯 (Rice)<br>レシピ (Recipe) | 森友学園(*Moritomo Gakuen*)<br>国会 (Congress)<br>自民党 (Liberal Democratic Party)<br>安倍 (Prime minister)<br>共謀罪 (Conspiracy law) | バンド (Band)<br>アーティスト (Artist)<br>アルバム (Album)<br>公演 (Performance)<br>不動産 (real estate) | 自民党 (Liberal Democratic Party)<br>脱原発 (Abandoning nuclear power)<br>維新 (Japan Restoration Party)<br>稼働 (Operation)<br>国会 (Congress) |
| 0: Abe | 0.23 | 0.15 | 0.01 | 0.04 | 0.00 | 0.04 | 0.00 |
| 1: Right-wing | 0.02 | 0.02 | 0.57 | 0.02 | 0.06 | 0.01 | 0.02 |
| 2: Renho | 0.10 | 0.05 | 0.01 | 0.17 | 0.01 | 0.06 | 0.02 |
| 3: Left-wing | 0.01 | 0.01 | 0.01 | 0.04 | 0.39 | 0.02 | 0.18 |
| 4: Yamamoto | 0.08 | 0.02 | 0.00 | 0.13 | 0.02 | 0.57 | 0.01 |
| 5: JRP and ruling party | 0.07 | 0.01 | 0.06 | 0.02 | 0.07 | 0.02 | 0.45 |

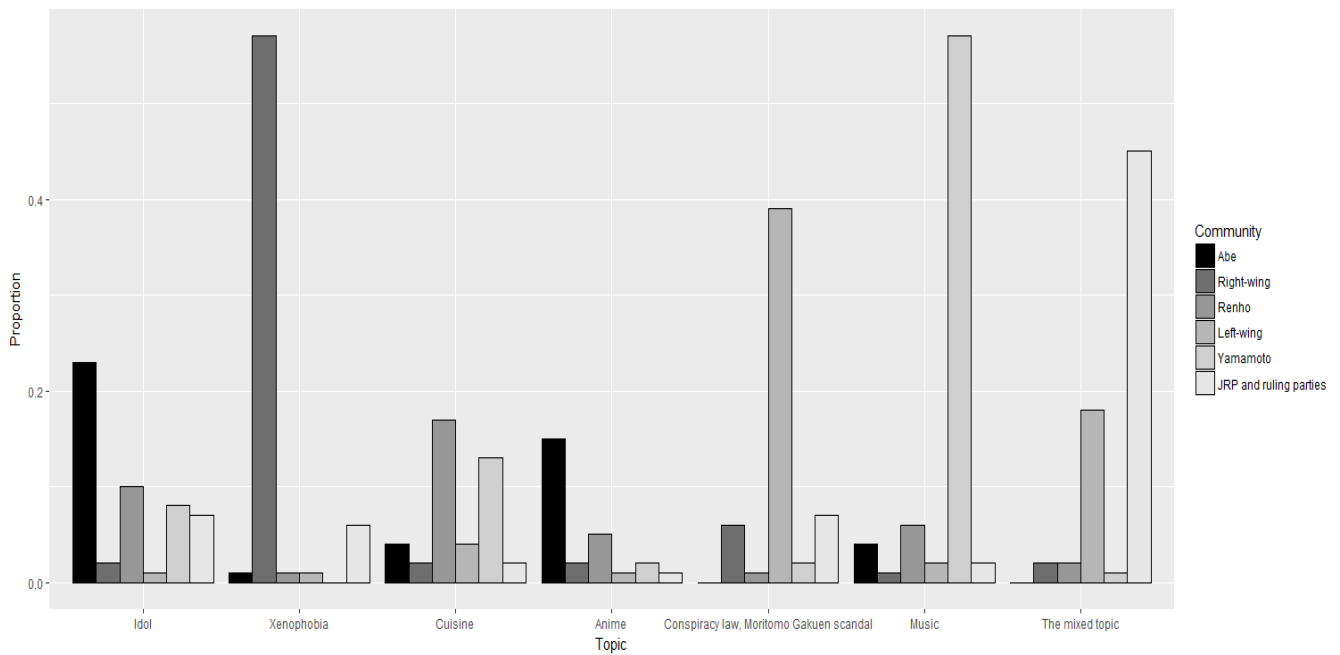

Figure 2. Distribution of Topics for Each Community

## VI. RESULTS

First, we applied Louvain methods to the analysis of the reciprocal following network (the number of nodes of which was 87,215 and the number of edges was 6,745,449). The network was partitioned into 113 distinct communities. We excluded the nodes that belong to the community (a) whose members are too few, (b) whose members are largely from foreign countries, or (c) whose members are bots[1]. This gave us six distinct communities.

Figure 1 shows the result of network visualization. (We only visualized the network of the users with more than 5000 followers. Visualization was done by Gephi.) The size of each community is shown in the second column of Table 1.

Next, in order to specify political orientation that characterized each community, we calculated the probability of following a political leader for each community member. Table I summarizes the results, which show that the patterns of following political leaders are differentiated among distinct communities. Community 0 largely consists of those who follow Shinzo Abe, the leader of Liberal Democratic Party, which is Japan's ruling party with conservative ideology. The ratio of Abe followers is 0.74. Community 1 is characterized as right-wing followers because more than half of the members follow the "Nippon no Kokoro" account, which is the rightist party in Japanese national politics. We will call this community right-wing followers community. Community 2 is characterized as Democratic Party in Japan (DPJ) followers, as the half of members follow Renho, who was the leader of DPJ at the time of data collection. Community 3 is left-wing followers, because members follow left-wing party leaders such as Mizuho Fukushima, the leader of Social Democratic Party, and Kazuo Shii, the leader of Japan Communist Party. The majority of Community 4 follow Taro Yamamoto, conjoint representative of Liberal Party. He is an activist for anti-nuclear power movement and a well-known actor in theater, cinema, and TV drama. We inferred from the profile description (the result is not shown) and the topic ratio below that members of Community 3 focus on his activity in the anti-nuclear power movement, whereas members of Community 4 are interested in him as a famous actor. Community 5 consists of followers of Ichiro Matsui, the leader of Japan Restoration Party, and Natsuo Yamaguchi, the leader of KOMEI party, which is the other ruling party at the time of data collection.

The next step was to investigate whether each community discusses different topics or shares a common interest across communities, by examining the overall topic distribution in each community.

We applied a topic model to our tweets contents corpus. We set the number of topics to be 50 for topic interpretability. We tried the different number of topics (30, 40, 60, 70, 80) and found the results were qualitatively similar. Some of the resulting topics are shown in Table II.

To answer the question of how differently distributed the topics are among distinct communities, we examined the topic proportion in each community and found that some speech communities are characterized by a few dominant topics.

Figure 2 shows the part of the result. The most salient is that members of the right-wing follower community use a set of words that express xenophobia (Topic 34) much more often than others. This topic accounts for 57% of the topics in the right-wing follower community but only for a maximum of 6% in other communities. A similar, but less extreme tendency is found in the left-wing follower community. The members in this community show great interest in conspiracy law and the Moritomo Gakuen scandal, which topics garner much less attention in other communities[2].

The Abe-following community and Renho-following community are characterized by a relatively even topic proportion and more mundane topics. Although the interest in Yamamoto community is exclusively focused on music, this should not be interpreted as the symptom of echo chambers. Rather, the members of this community may only have an interest in Yamamoto as an actor, not politician, thus making this community relatively distinct from other communities in this political field. Finally, JRP and the ruling-party follower community shows their interest in JRP related issues[3].

In summary, we found six major communities in the Twitter political field in Japan that are distinct from each other in the pattern of following political leaders. Topic analysis revealed topic polarization, especially among right-wing followers, whose speeches were characterized by xenophobia and might be echo-chambered within their community without exposure of counterarguments from other communities. The left-wing community also shows similar tendency, although the dialogue is more open than in the right-wing follower community.

## VII DISCUSSION

In this study, we tried to answer the question of whether there are distinct political speech communities in Japanese Twitter political field. Because there would be more than simply two camps of political ideologies (i.e., left and right), we adopted the inductive approach to see whether political distinct communities emerged in a bottom-up way through community detection rather than imposing predefined political categories. We found six distinct communities in the field that reflect a wide variety of political ideological positions in Japan. Very roughly speaking, Community 3 can be positioned on the leftmost side, Community 2 on the middle left, Community 0 on the middle right, and Community 1 on the rightmost side. Community 5 is difficult to position on the left–right scale. Community 4 may consist of less politically engaged people than others. Our inductive methods may be useful for Twitter political fields comprising multiple political positions that are difficult to be reduced to a simple left–right scheme.

---

[1] We disregarded the communities with less than 10 members. We judged whether the accounts are from foreign countries and whether they are bots from the contents of tweets and profile descriptions.

[2] This point may partly depend on the number of topics. For instance, when the number of topics is set to be 70, the issue related to conspiracy law and the Moritomo Gakuen scandal is differentiated into two different topics. In this case, one topic, Moritomo Gakuen scandal, is shared among different communities and the other, conspiracy law, is dominantly discussed about solely by the left-wing follower community. However, there remains the tendency that speech communities discussed about a few idiosyncratic issues which are not shared by other communities.

[3] Although the topic analysis with 50 topics failed to differentiate the content of the "mixed topic", word frequency analysis and topic analysis with the different number of topics revealed the following fact (results are not shown); JRP and the ruling-party community showed interest in JPP related issues such as Osaka metropolis plan while the nuclear plant problems are exclusively discussed by the left-wing follower community.

Furthermore, Community 1 is composed of around 10% of the total nodes, which is much larger than the estimated size of *Netto Uyoku* (0.1% in [24]). Of course, we limited our sample to moderate opinion leaders who are willing to share political information, exchange political opinions, and engage in political dialogue. Characteristics of our sample are not equal to *Netto Uyoku* even though writing about political issue on an online board is one feature of the group. However, at least when it comes to moderate opinion leaders, proportions of those in radical right community are not so small. This is a finding with serious implications because the moderate leaders in our sample all have more than 500 followers each, who read hate speech daily in their Twitter feed and may be influenced by these tweets.

Furthermore, we found that each community discusses different issues that rarely overlap. Right-wing followers write about Korea, Korean Japanese, and dual nationality issues. Left-wing followers write about conspiracy law and the corruption of the government. Because of their differing interests, there is little debate over these issues across community lines. Our results suggest that whether an echo chamber occurs in Twitter depends on the topic.

Our finding about the relationship between echo chambers and political ideologies is consistent with previous studies in spite of the difference in methodologies. Boutyline and Willer found that both more extreme and more conservative individuals are likely to form homophilic ties than more liberal and more moderate ones [5]. This finding is supported by the analysis of the formal structure of the Twitter field. Our finding complements this by examining the tweet contents, which analysis found that the extreme conservative community has the strongest tendency to circulate one exclusive topic within it; to a lesser degree, the leftist community also focuses on a few of topics more than the moderate ideology community.

Furthermore, our method contributes to the understanding of what kinds of topics are likely to be circulated within each speech community. Our result suggests that xenophobic speeches tend to circulate only within one idiosyncratic community. This is probably because the members in other communities hesitate to discuss or even refer to such topics. We should add that this does not justify the circulation of hate speech in Twitter because people can be exposed to such speeches even if they do not participate in the discussion or refer to them.

To our knowledge, this is the first study to address echo chamber issues in Japanese Twitter political field by examining the formal structure and the rich contents of tweets with the combination of large-scale social network analysis and natural language processing. We believe this method can be extended to the Twitter field or other social media analysis in other countries. In future work, it will be interesting to conduct an international comparison of the mechanism of political polarization with our method.

ACKNOWLEDGMENT

This work was supported by JSPS KAKENHI Grant Numbers 16K13406, 16K04027 and 16K13347. We thank Yoshimichi Sato, Hiroshi Hamada, and Takafumi Ito for helpful comments and advice on an earlier version of this paper.

REFERENCES

[1] J. Adams and V. J. Roscigno. J., 2005. "White supremacist, oppositional culture and the world wide web," Social Forces, vol. 84, 2005, pp. 759–778.

[2] P. Barberá, "Birds of the same feather tweet together: Bayesian ideal point estimation using Twitter data." Political Anal., vol. 23, 2014, pp. 76–91.

[3] P. Barberá, N. Wang, R. Bonneau, J. T. Jost, J. Nagler, J. Tucker, and S. González-Bailón. "The critical periphery in the growth of social protests." PloS One, vol. 10, 2015, p.e0143611.

[4] D. M. Blei, A. Y. Ng, and M. I. Jordan. "Latent dirichlet allocation." J. Mach. Learn. Research, vol. 3, 2003, pp. 993–1022.

[5] A. Boutyline and R. Willer "The social structure of political echo chambers: Variation in ideological homophily in online networks." Political Psychology, vo. 38, 2017, pp. 551–569.

[6] C. Budak and D. J. Watts. "Dissecting the spirit of Gezi: Influence vs. selection in the Occupy Gezi movement." Sociological Sci., vol. 2, 2015, pp. 370–397.

[7] M. Castells. "The new public sphere: Global civil society, communication networks, and global governance." The Annals of the American Academy of Political Social Sci., vol. 616, 2008, pp. 78–93.

[8] E. Colleoni, A. Rozza, and A. Arvidsson. "Echo chamber or public sphere? Predicting political orientation and measuring political homophily in Twitter using big data." J. Commun., vol. 64, 2014, pp. 317–332.

[9] M. Conover, J. Ratkiewicz, M. R. Francisco, B. Gonçalves, F. Menczer, and A. Flammini. "Political polarization on twitter." ICWSM, vol. 133, 2011, pp.89–96.

[10] P. Dahlgren. "The Internet, public spheres, and political communication: Dispersion and deliberation." Political Commun., vol. 22, 2005, pp. 147–162.

[11] W. A. Galston. "If political fragmentation is the problem, is the Intemet the solution?" In D. M. Anderson & M. Cornfield (Eds.), *The Civic Web: Online Politics and Democratic Values* (pp. 35^14). Lanham, MD: Rowman & Littlefield, 2003.

[12] Y. Halberstam and B. Knight. "Homophily, group size, and the diffusion of political information in social networks: Evidence from Twitter." J. Public Economics, vol. 143, 2016, pp. 73–88.

[13] N. Higuchi. *Nihongata Haigaishugi* (Japanese-style Xenophobia). Nagoya Daigaku Shuppankai, 2014.

[14] I. Himelboim, S. McCreery, and M. Smith. "Birds of a feather tweet together: Integrating network and content analyses to examine cross-

ideology exposure on Twitter." Journal of Computer‐Mediated Commun., vol. 18, 2013, pp. 40–60.
[15] L. Hong and B. D. Davison. Empirical study of topic modeling in twitter. In Proceedings of the first workshop on social media analytics, July 2010, pp. 80–88. ACM.
[16] A. Klein. "Slipping racism into the mainstream: A theory of information laundering." Commun. Theory, vol. 22, 2012, pp. 42–-448.
[17] H. Kwak, C. Lee, H. Park, and S. Moon. What is Twitter, a social network or a news media?.In *Proceedings of the 19th int. conf. on World wide web,* April 2010, (pp. 591-600). ACM.
[18] Ministry of International Affairs. (2015) "*Shakaikadai Kaiketsu no tameno Aratana ICT sarbisu/ gijutsu heno Hitobito no Ishiki ni Kansuru Chosa Houkokusho*" (Report on Public Opinions about New ICT Services and Technologies for Solving Social Issues). Ministry of International Affairs.
[19] Ministry of International Affairs. (2014) *Jouhou Tsushin Hakusho* (White Paper Infromation and Communications in Japan). Ministry of International Affairs.
[20] R. Mourtada and F. Salem. Civil movements: The impact of Facebook and Twitter. *Arab Social Media Report*, *1*(2), 2011, pp.1-30.
[21] C. R. Sunstein. # Republic: Divided Democracy in the Age of Social Media. 2017. Princeton University Press.
[22] F. Taka. *Reishizumu wo Kaibousuru* (Anatomy of Racism). 2015, Keiso Shobo.
[23] D. Tsuji. "Keiryou Chosa kara Miru 'Netto Uyoku' no Purofairu: 2007/2014nen Webuchosa no Bunsekikekka wo Motoni" (A Profile of Netto-uyoku: Quantitative Data Analysis of Online Questionnaire Surveys in 2007 and 2014). *Nenpou Ningen Kagaku*, 38, 2017, 211-224.
[24] D. Tsuji. *Intarnetto ni okeru 'Ukeika' Gensho ni kansuru Jissho Kenkyu Chousa Gaiyo Houkokusho* (Report on Empirical Research about "Rightward Trend" in the Internet), 2008, (http://d-tsuji.com/paper/r04/, last accessed 2017/09/27).
[25] D. Yasuda. *Netto to Aikoku* (Internet and Patriotism), 2012, Koudansha.